\documentclass[aps,unsortedaddress,nofootinbib]{revtex4}
\usepackage{amsmath}
\begin{document}

\title{Approaching  Space  Time  Through  Velocity  in  Doubly  Special
  Relativity}

\author{R.~Aloisio}
\affiliation{INFN, Laboratori Nazionali del Gran Sasso,\\
67010 Assergi, L'Aquila, Italy}
\email{aloisio@lngs.infn.it}

\author{A.~Galante}
\affiliation{Dipartimento di Fisica dell'Universit\`a di L'Aquila,\\
67100 L'Aquila, Italy}
\affiliation{INFN, Laboratori Nazionali del Gran Sasso,\\
67010 Assergi, L'Aquila, Italy}
\email{galante@lngs.infn.it}

\author{A.F.~Grillo}
\affiliation{INFN, Laboratori Nazionali del Gran Sasso, \\
67010 Assergi, L'Aquila, Italy}
\email{grillo@lngs.infn.it}

\author{E. Luzio}
\affiliation{Dipartimento di Fisica dell'Universit\`a di L'Aquila,\\
67100 L'Aquila, Italy}

\author{F.~M\'endez}
\affiliation{INFN, Laboratori Nazionali del Gran Sasso,\\ 
67010 Assergi, L'Aquila, Italy}
\email{mendez@lngs.infn.it}

\begin{abstract}
We  discuss  the definition  of  velocity  as  $dE/d|{\bf p}|$,  where
$E,{\bf  p}$ are  the energy  and momentum  of a  particle,  in Doubly
Special  Relativity  (DSR).   If  this  definition  matches  $dx/dt  $
appropriate  for  the  space-time   sector,  then  space-time  can  in
principle  be built  consistently with  the existence  of  an invariant
length scale.  We show  that, within different  possible velocity
definitions, a space-time compatible with momentum-space DSR principles 
can not be derived.

\end{abstract}

%\keywords{Models of Quantum Gravity, Space-Time Symmetries}
\maketitle

\newpage

\section{Introduction}
Doubly Special  Relativity (DSR) is  a generalization of the 
special relativity which, besides of the speed of light $c$, introduces a
second  invariant scale  \cite{ame1,ame2,ame3,ms,alu}.  The  reason to
look for such generalization can be traced back to an heuristic argument
of  quantum  gravity  (QG):  if  quantum  effects  of  gravity  become
important  at  certain  distances  (typically the  Planck  length)  or
energies, then these scales should be observer independent.

With this in mind, the DSR proposal tries to give a (at least
phenomenological)  answer  to the  question  raised  by the  previous
argument, that is, if it is possible to find a different symmetry that
guarantees another  invariant scale (which will  be eventually related
with the standard QG scale).  A concrete realization of such ideas was
given in the  space of energy and momentum  where the deformed boosts,
dispersion relation and composition law, were written \cite{ms,alu,techdsr}.

DSR can also be understood as a non-linear realization of the Lorentz 
group in the momentum
space   \cite{ms,nldsr,lukdsr}  $i.e.$,   apart   from  the   physical
variables,  we  can  consider  auxiliary  variables
that define a space where the Lorentz group acts linearly
\footnote{
  Since the auxiliary variables transform according to the standard
       Lorentz group, we will refer to them and to the space 
       where they are defined as {\em Classical Variables} and {\em 
       Classical Momentum Space} respectively.
 }.  
However the program  has not be completed yet.  One of
the  most pressing problems  on this  subject is  to find  an explicit
realization  of these  principles  in  the space  time.  In fact,  the
connection between  the existence of  an energy or momentum  (or both)
invariant scale and the consequences in the physical space-time is not
clear.  This is a necessary step  to undergo since it is in the actual
space-time sector where the experiments are performed, the instruments
collect data and, finally, our physical description has to apply.

One possible way  to approach this problem is by  noticing that in the
usual case  (relativistic and non relativistic)  there are quantities,
like the velocity of a particle,  that can also be written in terms of
variables of the energy-momentum space. In the standard description, 
the velocity
is the derivative  of the spatial coordinate with  respect to time and
it is also the derivative of the energy with respect to momentum.

In this work, we analyze  two possible definitions for the velocity of
a particle and we test  their implications for the space time (problem
treated for  first time in  \cite{amelino2}).  The first  case studied
corresponds  to the  standard definition  mentioned above  which, when
physical processes are considered,  gives rise to inconsistencies (also
reported in \cite{mignemi,deg}).

As a different possibility, we have used the map that
connects the momentum space with the {\em classical momentum space} and,
since  both  approaches  give  rise  to  the  same  expression  for
velocities (see  also the discussion in \cite{gran}), we conclude that
attempts to  define the space time of  DSR in terms of  a {\em classical
space time} shall give inconsistent results.

Another  possible definition of  velocities analyzed  in this  work is
related to  a deformation  of the definition  of derivatives.  In this
approach we adopt  again the notion of velocity as  the rate of change
of the energy  with respect to the change of  momentum, but the notion
of  {\em  change} adopted is now DSR  compatible.  That  is,  the
difference (for energy and also for momenta) is covariant under a DSR
boost. This   definition   of   velocity  does   not   show   the
inconsistencies described before while it is still connected to a limited
three-momentum; however  also in this case it  seems inconsistent with
a continuous differential space time manifold. 

The  latter   approach  is  close   to  the  velocity   definition  in
$\kappa$-Poincar\'e      (KP)     and      $\kappa$-Minkowski     (KM)
\cite{lukvel,taka,who} and to  the DSR approach where a  KM space time
is assumed \cite{kowvel,kowvel2}.

It is important however to note that  the definitions of {\em velocity}
we will describe in detail in the  rest of the paper are formal and it
is not clear if and how they are related to the
rate of  change of space with  time and so  are their phenomenological
implications. 

This work  is connected  to some  early efforts in  the approach  to a
space-time formulation  of DSR.  Following the  close relation between
DSR and the algebraic sector of the so called $\kappa$-Poincar\'e (KP)
deformation  of  the Poincar\'e  group  \cite{KP},  some authors  have
introduced the  idea that  the DSR compatible  space-time should  be a
non-commutative  space-time,   as  it  happens   with  the  space-time
associated  to KP  \cite{KPDSR}. Following  the approaches  similar to
DSR, other authors  have investigated possible non-linear realizations
of the Lorentz group, directly on the space-time \cite{dsrsptm}.

The paper is  organized as follow. In the next  section we will review
the formulation of DSR as a nonlinear realization of the Lorentz group
in the  momentum space. In  section three, the standard  definition of
velocities will be reviewed and  analyzed. Section four is devoted to
the analysis in the so called {\em classical space}. The definition of
velocities with a DSR inspired derivative is given in Section five. The
last section is devoted to the discussion and conclusions. 

%%%%%%%%%%%%%%%%%%%%%%%%%%%%%%%%%%%%%%%%%%%%%%%%%%%%%%%%%%%%%%%%%%%%%%
%%%%%%%%%%%%%%%%%%%%%%%%%%%%%%%%%%%%%%%%%%%%%%%%%%%%%%%%%%%%%%%%%%%%%%
\section{DSR formulation}
%%%%%%%%%%%%%%%%%%%%%%%%%%%%%%%%%%%%%%%%%%%%%%%%%%%%%%%%%%%%%%%%%%%%%%
%%%%%%%%%%%%%%%%%%%%%%%%%%%%%%%%%%%%%%%%%%%%%%%%%%%%%%%%%%%%%%%%%%%%%%
Considering    the     extensive    literature    on     this    topic
\cite{ame1,ame2,ame3,ms,alu,techdsr,lukdsr,nldsr,amelino2,more1,more2},     we
only briefly  summarize what we  will define as a  DSR.  Understanding
DSR as  a non linear realization  of Lorentz symmetry  we consider two
momentum spaces: one, that we call {\em the classical momentum space} $\Pi$, with
coordinates   $\pi_\mu=\{\epsilon,{\boldsymbol   \pi}\}$,  where   the
Lorentz group acts linearly, and another, the physical space $P$, with
coordinates  $p_\mu=\{E,{\boldsymbol p}\}$.   There exists  a function
$F:P\rightarrow  \Pi$, such  that  $\pi =  F[p]$\footnote{We omit  the
indexes  in  order  to  simplify  the  notation.}.   This  function  is
invertible and  depends on  a parameter $\lambda$.   The image  of the
point  $p_*=(E,{\boldsymbol p}_{max})$ (where  ${\boldsymbol p}_{max}$
is  a vector with  modulus $|{\boldsymbol  p}_{max}|=1/\lambda$) under
$F$  is infinity;  this  requirement ensures  that  $1/\lambda$ is  an
invariant momentum scale.

Boost transformations and the Casimir  elements in $P$ are the inverse
images of the boosts and Casimir in the classical space.  That is, the
boost in the $P$ space is given by
\begin{equation}
\label{dsrboost}
{\cal B}=F^{-1}\circ\Lambda\circ F,
\end{equation}
where $\Lambda$ is an element of the Lorentz group. 

The Casimir in the classical space is $\epsilon^2 - {\boldsymbol
  \pi}^2=\mu^2$, which can be written in the $P$-space as $F_0[p]^2 -
 { \boldsymbol F}[p]^2=\mu^2$. 

For the DSR1 model \footnote{In  the following we will always refer to
  this specific realization.}, the explicit form of $F$ and its inverse
  is \cite{lukdsr}
\begin{eqnarray}
\label{f}
F^{-1}[x,\boldsymbol{y}]&=&\left[\begin{array}{c}
\frac{1}{\lambda}\ln\left(\lambda
x+\sqrt{1+\lambda^2(x^2-\boldsymbol{y}^2)}\right)   \\   
\boldsymbol{y}\left[\lambda
x+\sqrt{1+\lambda^2(x^2-\boldsymbol{y}^2)}\right]^{-1}
\end{array}\right],
\\                           
F[x,\boldsymbol{y}]&=&\left[\begin{array}{c}
\frac{1}{\lambda}\left(\sinh(\lambda    x)+\frac{\lambda^2}{2}\boldsymbol{y}^2
e^{\lambda x}\right) \\ \boldsymbol {y}e^{\lambda x}
\end{array}\right].
\end{eqnarray}

It is then  possible to write the explicit formulas for the boosts and
the  Casimir elements.  Because  of the  rotation  invariance we  will
always reduce the problem to  1+1
dimensions so, without loss of generality, we can write the boost for
energy and momentum 
\begin{eqnarray}
\label{expboost}
E(\xi)&=&           E+\frac{1}{\lambda}\ln\left[1+\lambda          p_x
\sinh(\xi)-\left(1-\cosh(\xi)\right)\left(\sinh(\lambda   E)e^{-\lambda
E}+\frac{\lambda^2   {\boldsymbol   p}^2}{2}\right)\right],   \\   p_x(\xi)&=&
\frac{p_x \cosh(\xi) + \sinh(\xi)\left(\left(\lambda^{-1}\sinh(\lambda
E)e^{-\lambda  E}+\frac{\lambda {\boldsymbol p}^2}{2}\right)\right)}{1+\lambda
p_x             \sinh(\xi)-\left(1-\cosh(\xi)\right)\left(\sinh(\lambda
E)e^{-\lambda    E}+\frac{\lambda^2    {\boldsymbol    p}^2}{2}\right)}.
\end{eqnarray}
Here  $\xi$  parametrizes  the  elements  of the  Lorentz  group  and
therefore $-\infty < \xi <\infty$.  
 
The Casimir turns out to be
\begin{equation}
\label{casdsr}
\cosh(\lambda E) -\frac{\lambda^2 {\boldsymbol p}^2}{2}e^{\lambda
  E}=\cosh(\lambda m),
\end{equation}
with $m=m(\mu)$ the physical mass, that is, it coincides with the mass of the
particle in  the limit  $\lambda \rightarrow 0$.

Finally, in order  to have a full description of the  way in which the
measurements made by one observer are related to the measurements made
by another observer boosted with respect to the first, it is necessary
to  know  the  relation  between,  $\xi$  and  the  relative  velocity
$V$. This  will allow  to check the  consistency of the  definition of
velocities, a quantity that in principle can be measured by
experiments.

%%%%%%%%%%%%%%%%%%%%%%%%%%%%%%%%%%%%%%%%%%%%%%%%%%%%%%%%%%%%%%%%%%%%%%
%%%%%%%%%%%%%%%%%%%%%%%%%%%%%%%%%%%%%%%%%%%%%%%%%%%%%%%%%%%%%%%%%%%%%%
\section{On the Definition of relative velocity}
%%%%%%%%%%%%%%%%%%%%%%%%%%%%%%%%%%%%%%%%%%%%%%%%%%%%%%%%%%%%%%%%%%%%%%
%%%%%%%%%%%%%%%%%%%%%%%%%%%%%%%%%%%%%%%%%%%%%%%%%%%%%%%%%%%%%%%%%%%%%%
Our  final   aim  would  be   to  investigate  the   relation  between
measurements made by two
observers in  relative motion and,  through that, the structure  of the
space time. To complete this  program it is first necessary to discuss
the  definition  of  velocities   and  its  relation  with  the  boost
parameter.

This section is devoted to  investigate such topics. We will adopt the
standard definition of velocity \cite{ame1} and study
its consequences  from the point of  view of measurements  made by two
inertial observers.

%%%%%%%%%%%%%%%%%%%%%%%%%%%%%%%%%%%%%%%%%%%%%%%%%%%%%%%%%%%%%%%%%%%%%%
\subsection{Definition of velocity and its relation with $\xi$}
%%%%%%%%%%%%%%%%%%%%%%%%%%%%%%%%%%%%%%%%%%%%%%%%%%%%%%%%%%%%%%%%%%%%%%
The physical scenario consists of two observers in relative uniform
motion.  $S'$ is the reference frame where we consider particles  at rest
(it  corresponds to  $\xi=0$), while  $S$ is  another  reference frame
whose  motion  respect  to $S'$  is  described  by  a non  zero  boost
parameter $\xi$.

Since at present stage we only know the DSR transformation laws in the
energy  momentum sector, we  will consider  only measurement  of these
quantities.  Clearly what we do not know are the transformation laws in
the space  time sector and, in  fact, nothing guarantees  us even that
the space-time is a continuous and differentiable manifold.

In undeformed relativity  (and also in the Galilean  case) it is possible
to express the velocity in terms of the momentum and the energy.  This
is provided by the relation
\begin{equation}
\label{veloc}
V=\frac{dE}{d p}, 
\end{equation}
where $E$ is  the energy of the particle, $  V =|{\boldsymbol v}|$ and
$p=|{\boldsymbol p}|$.

This expression, that  gives the right results for  the standard cases
mentioned above, might not be correct in our case.  One must note that
(\ref{veloc})  is  based  on  the  facts  that  $a)$  there  exist  an
expression for  the energy in  terms of coordinates and  momenta (in
general, one  assumes that  is possible to  define a  Hamiltonian that
generates time translations) and $b)$ there exist a canonical simplectic
structure \cite{velsimp}. 

Both ingredients  are independent and, regarding the  point $a)$, here
we  assume  that  the  energy  in  terms  of  momenta  is  given  by
(\ref{casdsr}).  Note also that  if we  follow the  standard approach,
what  we  have called  the  velocity  should  be identified  with  the
variation of coordinates with respect to  time: this is the outcome
of the Hamilton equations. 

Starting  from  the  above  working hypothesis,  from  the  dispersion
relation (\ref{casdsr}) we can calculate the velocity previously defined
\begin{equation}
\label{vp}
V(p)=\frac{\lambda~ p}{r^2(\lambda p) }\left(1+\frac{{\cal C}_{\lambda
    m}^2(\lambda p)}{{\cal  C}_{\lambda m}^2(\lambda p)  - r^2(\lambda
    p)} \right),
\end{equation}
with   $r^2(\lambda  p)=1-p^2   \lambda^2$   and  ${\cal   C}_{\lambda
m}^2(\lambda            p)=\cosh^2(\lambda            m)+\cosh(\lambda
m)\sqrt{\cosh^2(\lambda m) - r^2(\lambda p)}$. 

Now let us assume that there is a particle of mass $m$ at rest in $S'$ 
(where {\em rest} means that ${\boldsymbol p}=0$)
and we observe it from $S$.  Since we
can relate momenta measured in $S$ with the same quantities measured
in $S'$ by using (\ref{expboost}), we are able to express the velocity
of the  particle in  terms of  its mass and  the parameter  $\xi$.  We
obtain
\begin{equation}
\label{vxi}
V(\xi)=   \frac{\sqrt{1-r^2_{\lambda  m}(\xi)}}{r_{\lambda  m}^2(\xi)}
 \left(1+\frac{{\cal    C}_{\lambda    m}^2(\xi)}{{\cal    C}_{\lambda
 m}^2(\xi) - r_{\lambda m}^2(\xi)} \right),
\end{equation}
with
$$             r^2_{\lambda            m}(\xi)=1-\frac{\sinh^2(\lambda
m)~\sinh^2(\xi)}{\left(\cosh(\lambda    m)    +    \sinh(\lambda    m)
\cosh(\xi)\right)^2},
$$
$$   {\cal   C}_{\lambda  m}^2(\xi)=\cosh^2(\lambda   m)+\cosh(\lambda
m)\sqrt{\cosh^2(\lambda m) - r_{\lambda m}^2(\xi)}.
$$

At first sight, (\ref{vxi}) seems to have a pole for ${\cal
C}^2_{\lambda  m}(\xi)  =  r^2_{\lambda  m}(\xi)$,  which  occurs  for
$m=0$. This is not true and  indeed the $m\to 0$ limit is well defined
and gives  $V(\xi)=\tanh(\xi)$.  Notice that this is  not the velocity
of  a photon  since it  corresponds  to the  limit $p_\mu^{'}=0$,  and
therefore describes the motion of a geometrical point. 

A fundamental property of (\ref{vxi}) is its mass dependence (see also
discussions in  \cite{mignemi,deg}). In order to  discuss the consequences
of the definition, as well as  the meaning of this mass dependence, in
the next subsection we will analyze some limits and peculiarities of
the previous expression.

%%%%%%%%%%%%%%%%%%%%%%%%%%%%%%%%%%%%%%%%%%%%%%%%%%%%%%%%%%%%%%%%%%%%%%%%%%%%%
\subsection{Special limits and the mass dependence of the relative velocity}
%%%%%%%%%%%%%%%%%%%%%%%%%%%%%%%%%%%%%%%%%%%%%%%%%%%%%%%%%%%%%%%%%%%%%%%%%%%%%

In (\ref{vxi}) we see that  the mass always appears in the combination
$\lambda  m$.  Since  $\lambda$   is  the  parameter  controlling  the
departure from standard Lorentz invariance  it is interesting
to consider the large mass limit  ($\lambda m >>1$), which we call the
{\em macroscopic bodies limit}, and the limit of small masses $\lambda
m <<1$ that will be referred as {\em microscopic bodies limit}.

%%%%%%%%%%%%%%%%%%%%%%%%%%%%%%%%%%%%%%%%%%%%%%%%%%%%%%%%%%%%%%%%
\subsubsection{Microscopic bodies limit ($\lambda m<<1$)}
%%%%%%%%%%%%%%%%%%%%%%%%%%%%%%%%%%%%%%%%%%%%%%%%%%%%%%%%%%%%%%%%
It is expected that for  particles with masses far below the invariant
scale, the relativistic  limit must be recovered. It  is not difficult
to show that, for any value of $\xi$
\begin{equation}
\label{velpicc}
V_{\lambda m<<1}\sim \tanh(\xi)\left[1+ \tanh(\xi)\sinh(\xi)~ \lambda
  m -\frac{1}{2}\tanh^2(\xi)~( \lambda m )^2+\cdots\right]. 
\end{equation}
We see that  the zero order term corresponds  to the relativistic case.

It is  also interesting to note  that, in this  microscopic limit, the
momentum of the particle seen by $S$, is given by
\begin{equation}
\label{macromom}
p_x\sim       m~\sinh(\xi)-\lambda       m^2\cosh(\xi)\sinh(\xi)+{\cal
  O}(\lambda^2),
\end{equation}
and we see again that the first term is the standard relativistic one.

Finally, let us note that since all the relativistic limits are
recovered for microscopic bodies, if we consider now $\xi
\rightarrow 0$ we reproduce the Galilean limit for momentum as well as
for the relative velocity.

%%%%%%%%%%%%%%%%%%%%%%%%%%%%%%%%%%%%%%%%%%%%%%%%%%%%%%%%%%%%%%%
\subsubsection{Macroscopic bodies limit ($\lambda m>>1$)}
%%%%%%%%%%%%%%%%%%%%%%%%%%%%%%%%%%%%%%%%%%%%%%%%%%%%%%%%%%%%%%%
Following the previous analysis, it is natural to consider the limit
$\lambda  m  >>1$:  this  limit  is not  forbidden  because,  in  DSR,
particles have a  maximum momentum attainable but the  energy (and the
mass) can be as large as we want.

The velocity in this limit becomes
\begin{equation}
\label{velgran}
V_{\lambda m>>1}\sim \sinh(\xi)+2~e^{-2 \lambda m} \tanh(\xi)+\cdots.
\end{equation}
>From here  it is clear that  the relativistic limit  is not recovered,
instead, for $\xi \to 0$ we recover the Galilean one. 

In order to understand this result,  we note that in DSR1 there exists
a  maximum  momentum   ($p_{max}=1/\lambda$)  which  is  an  invariant
dimensional scale.  Therefore the condition $\lambda m>>1$ means $m >>
p_{max}$,   something  that  resembles   what  occurs   in  undeformed
relativity  when  the transition  from  the  relativistic  to the  non
relativistic regime is considered.
In this sense, we can  expect the relation (\ref{velgran}) between $V$
and $\xi$ to be Galilean-like in the limit $\xi\to 0$ as indeed happens. 

We  could expect  a similar  behavior to  hold also  for  the momentum
variable.  The macroscopic limit for the momentum is given by
\begin{equation}
\label{macromo}
p = \frac{\sinh(\xi)}{\lambda(1+\cosh(\xi))}+ {\cal O}(e^{-\lambda m}).
\end{equation}
For large values of $\xi$ the momentum correctly goes to $p_{max}$. 

As a function of the velocity, the previous expression turn out to be
\begin{equation}
\label{macromogal}
p=\frac{V}{\lambda(1+\sqrt{1+V^2})} + {\cal O}(e^{-\lambda m}),
\end{equation}
for any value of $\xi$.  In particular, when $\xi \to 0$ ---and
therefore the velocity (\ref{velgran}) is the Galilean one--- we see
that 
\begin{equation}
\label{macromogal2}
p\neq m ~V.
\end{equation}

Then, the relations we derived for $\lambda m>>1$ are perfectly acceptable
from  a DSR1  point of  view, but  they are  inconsistent  with everyday
experience for macroscopic bodies. 

This is related to the   so  called {\em  soccer ball  problem},
according to which, it would  be impossible to have macroscopic bodies
with  momentum  grater than  the  Planck's  momentum  (if we  identify
$1/\lambda = p_{pl}$) since all the particles in this body have a
limited momentum  and the DSR1  compatible composition law  allows only 
bodies with momenta no grater than $p_{max}$.

The problem  resides in  the fact that  this result is  not consistent
with our everyday experience, and the  reason is that we are trying to
use  DSR principles  (which  we expect  to  be relevant  as a  quantum
gravity effect) in the opposite extreme limit (the macroscopic one) where
quantum effects and especially quantum gravity effects are expected to
be  irrelevant.  A  true  space  time description  should  cure  these
apparent inconsistencies.

%%%%%%%%%%%%%%%%%%%%%%%%%%%%%%%%%%%%%%%%%%%%%%%%%%%%%%%%%%%%%%%%%%%%%%%
\subsubsection{The mass dependence of $V$}
%%%%%%%%%%%%%%%%%%%%%%%%%%%%%%%%%%%%%%%%%%%%%%%%%%%%%%%%%%%%%%%%%%%%%%%

In the standard Lorentz theory the velocity, defined as the derivative
of the energy respect to the momentum, is a function of
the boost  parameter without  any dependence on  the particle  mass or
energy.  Since the  boost  parameter has  a geometrical  significance,
irrespective on  the particle content  of the system,  this guarantees
that all  particle with  the same velocity  in a reference  frame will
have the same velocity in any other reference frame.

Now we try to give a physical interpretation to expression (\ref{vxi})
which,  comparing  to  the  standard  Lorentz case,  has  the  evident
peculiarity to define a mass dependent velocity. This can be stated in
a different  way by noting  that, inverting the  relation (\ref{vxi}),
the parameter of the boost depends on the mass of the particle.

Can  this result be  accepted?  To  answer this,  let us  consider two
observers, one in $S'$ and the  other in $S$ and two particles at rest
in $S'$ with different masses $m_1$ and $m_2$; a given value for $\xi$ describes 
the relative motion  of the observers. In this  framework the observer
in $S$  will measure  two different velocities  for the  two particles
and, consequently, there could be events observed by $S$ but that do
not happen  according to $S'$.   As an example  if we imagine  the two
particles at rest  at a given distance in $S'$,  since $m_1$ and $m_2$
have different  velocities in  $S$, it might  be possible to  observe a
collision according to $S$, something which will never occur according
to $S'$.  In  the previous argument, it is implicit  the fact that the
relation between velocity and the space time coordinates are the usual
one (in  the sense that two  particle at a finite  distance and finite
relative velocity  along a  given axes will  collide in a  finite time
interval).

In other words,  if we try to establish  the relative velocity between
two  reference  frames  measuring   the  velocities  of  particles  of
different mass at rest in one  of the reference frames we will clearly
get  different values.  To  better understand  the physical  origin of
this  apparently inconsistent  result we  can consider  again  our two
particles system. The total energy and momentum of the system is known
through  the  composition laws  which  are  compatible  with the  DSR1
principles.  DSR1 composition laws, at first order in $\lambda$ are
\begin{eqnarray}
\label{compodsr}
E_{tot}&\sim &E_1 + E_2 -\lambda p_1 p_2,
\\
p_{tot}&\sim &p_1 + p_2 -\lambda(E_1 p_2 + E_2 p_1),
\end{eqnarray}
which  can  be expressed  in  terms of  the  rapidity  $\xi$ by  using
(\ref{expboost}) at first order. 
\begin{eqnarray}
\label{compodsrxi}
E_{tot}(\xi)&\sim &M \cosh(\xi) - \frac{\lambda M^2}{2}\sinh^2(\xi),
\\
p_{tot}(\xi)&\sim &M\sinh(\xi)-\lambda M^2\sinh(\xi)\cosh(\xi),
\end{eqnarray}
with $M=m_1+m_2$. From here, using (\ref{veloc}), 
the velocity of the two particle system ($1+2$) is
\begin{equation}
\label{vtot}
V_{(1+2)}\sim \tanh(\xi)\left[1+ \lambda M ~\tanh(\xi)\sinh(\xi) +\cdots\right]. 
\end{equation}

This expression coincides  with (\ref{velpicc}) after the substitution
$m\rightarrow M$:  this is a  necessary consistency check  between the
definition of  velocity and the momentum composition  laws.  The point
is that the presence of more  than one particle turns out to introduce
problems in interpreting experimental  results.  If the particles have
different  masses or  if we  deal with  multiparticle systems,  we get
different  results   for  relative  velocities   of  reference  system
corresponding to a given boost parameter.

This    resembles    the   spectator    problem    that   arises    in
$\kappa$-Poincar\'e  models \cite{KP}  when we  consider  the deformed
composition law for  four momentum.  In that case  the composition law
is asymmetric under  the interchange of particles and  a particle that
does not participate in a  reaction process (the spectator) can modify
(simply by its presence, without any direct interaction) the threshold
energy for the process \cite{more2}.

What is remarkable is that in $\kappa$-Poincar\'e this problem has its
origin in  the asymmetry of the  momentum composition law  and, at the
end, this property  give rise to the non  commutative structure of the
space time \cite{NCKP}.  Here we deal with a symmetric composition law
of  momenta but,  assuming (\ref{veloc}),  we find  some inconsistency
similar   to  the  $\kappa$-Poincar\'e   spectator  problem.   With  a
speculative  attitude it  could be  interpreted as  a signal  that the
space time might have a non standard structure.

The above discussion points out the impossibility to physically accept
these results. The inconsistencies are  based on the assumption that a
unique  boost  parameter is  associated  to  a transformation  between
different  reference frames,  independent on  their  particle content.
This still leaves an open door: let us impose the physical requirement
that the velocities  measured by $S$ are equal  when the corresponding
particles are  at rest in  $S'$.  The above condition  reads $V_1=V_2$
and if  $m_1\ne m_2$ this necessarily implies  that the transformation
parameter has to be different for particles of different mass $i.e.$
\begin{equation}
V(\xi_1,\lambda m_1)=V(\xi_2,\lambda m_2).
\label{prrrr}
\end{equation}

The above formula relates the $\xi$ parameter
for  particles of  different  mass.  For example,  at  first order  in
$\lambda m$, in the case of two particles, we have 
\begin{equation}
\xi_2  =  \xi_1  +  \lambda  \sinh^3(\xi_1)(m_1  -  m_2)  +  {\cal  O}
(\lambda^2 (m_1-m_2)^2).  
\end{equation}
As a particular case we can express the $\xi_i$ of the $i$-particle as
a function of its mass and a boost parameter ($\xi$) that
corresponds to the $m \to 0, p \to 0$ limit discussed at the end 
of section 3.1.  This will allow us to write
\begin{equation}
\xi_i = \xi - \lambda m_i\sinh^3(\xi)  + {\cal O}
(\lambda^2 m_i^2),
\end{equation}
$i.e.$  to explicitly  express all  the mass  dependence of  the boost
parameter for any particle.  In this picture the boost parameter turns
out  to  be  particle   (mass)  dependent  under  a  precise  physical
requirement:  different  particles  will  have the  same  velocity  in
another reference frame if they are at rest in a given frame.

We  can  expect  this  to  turn into  contradiction  when  considering
multi-particle sates.  In fact the DSR four  momentum composition laws
are formulated to be covariant  under boosts when the {\it same} boost
parameter is  considered for each  particle.  The latter  condition is
not any more  satisfied if we assume (\ref{prrrr})  as can be verified
in 
the next example.   Consider two particles labeled 1 and  2 at rest in
the reference frame  $S'$. At first order in  $\lambda$ their momentum
in the  $S$ reference frame can be  written in terms of  the zero mass
boost parameter $\xi$ and the particle masses
\begin{eqnarray}
\label{PE}
p_i(\xi)&=&m_i\sinh(\xi)(1-\lambda m_i\cosh^3(\xi))+ {\cal O}(\lambda^2),\\
E_i(\xi)&=&m_i\cosh(\xi)(1-\lambda m_i 
\cosh(\xi)\sinh^2(\xi)(1+\tanh^2(\xi)))+ {\cal O}(\lambda^2). 
\end{eqnarray}

Now  we  consider this  two  particles as  a  unique  system.  In  the
reference frame  $S'$ we use  the composition law  (\ref{compodsr}) to
get   that  the   total  momentum   is   zero  and   the  energy   is
$M=m_1+m_2$. Then we can go to the reference system $S$ and write that
the total  momentum of the system  is simply given  by (\ref{PE}) with
$m_i$ replaced by the total mass $M$. 

We can  redo the calculations  first considering the momentum  of each
particle  in the  $S$ reference  system and  then composing  it (again
using the composition law (\ref{compodsr})) to get the total momentum.
For  consistency the latter  should coincide  with the  total momentum
previously obtained.  This is not the case. Composing the momentum in
$S$ we get
\begin{eqnarray}
\label{PE2}
P_{total}(\xi)&=&M\sinh(\xi)(1-\lambda M\cosh^3(\xi))+2\lambda m_1 m_2
\sinh^3(\xi)\cosh(\xi)+ {\cal O}(\lambda^2),\nonumber
\\
E_{total}(\xi)&=&M\cosh(\xi)(1-\lambda M 
\cosh(\xi)\sinh^2(\xi)(1+\tanh^2(\xi)))+\nonumber
\\
&&2\lambda    m_1   m_2\sinh^2(\xi)\cosh^2(\xi)(3+\tanh^2(\xi))+   {\cal
  O}(\lambda^2).\nonumber 
\end{eqnarray}
which differ  from the  previous result  due to the  last term  in the
r.h.s. of both equations. 

We conclude that both possibilities  $i.e.$ to consider a unique boost
parameter for all particles (like in the standard Lorentz formulation)
as well  as to  change it  to keep velocities  equal in  all reference
systems  for particles at  rest in  a given  reference frame,  are not
compatible with the velocity definition we adopted at the beginning of
this section.

%%%%%%%%%%%%%%%%%%%%%%%%%%%%%%%%%%%%%%%%%%%%%%%%%%%%%%%%%%%%%%%%%%%%%%
%%%%%%%%%%%%%%%%%%%%%%%%%%%%%%%%%%%%%%%%%%%%%%%%%%%%%%%%%%%%%%%%%%%%%%
\section{Velocity in the classical space}
%%%%%%%%%%%%%%%%%%%%%%%%%%%%%%%%%%%%%%%%%%%%%%%%%%%%%%%%%%%%%%%%%%%%%%
%%%%%%%%%%%%%%%%%%%%%%%%%%%%%%%%%%%%%%%%%%%%%%%%%%%%%%%%%%%%%%%%%%%%%%
Another way to obtain the relation between $V$ and the boost parameter
$\xi$ and  also to  get information about  the structure of  the space
time, is to study the  relation between quantities defined in the real
space and in the classical space. 

The velocity in the classical space $\Pi$, is naturally defined through
(\ref{veloc}), but in terms of the classical coordinates
\begin{equation}
\label{virveloc}
\nu = \frac{d \epsilon}{d\pi}.
\end{equation}

Then, it is possible to write the velocity (\ref{veloc}) in the real space, in
terms of variables in the classical space by using that $p=F[\pi]$,
with $F$ defined in (\ref{f}) and $\tanh(\xi) = \nu$.  Then 
\begin{equation}
\label{vivel}
V=\frac{\nu~\partial_\epsilon E + \partial_\pi E}{\nu~\partial_\epsilon p 
+ \partial_\pi p}.
\end{equation}

This velocity  trivially coincides  with (\ref{vxi}) because,  at this
level, we have only made a change of variables. 

What is  interesting to note is  that this result  suggests that there
could be also a map between a {\em classical  space time} and the real
space time.  In fact,  let us  assume that this  map exists.  That is,
there  exists  a  classical  space  time  $  \Omega$  with  coordinates
$\chi,\tau$ and two  functions $A[\chi,\tau], B[\chi,\tau]$ which also
depend  on the  parameter $\lambda$  (they  are the  analogous of  the
components of the function $F^{-1}$ in the momentum space), such that
\begin{equation}
x=A[\chi,\tau],\,\,\,\,\,\,\,\,\,\,\,\,\,\,\,t=B[\chi,\tau].
\end{equation}

In this case, posing $\nu =d\chi/d\tau$, the relation between the 
velocity in the classical space and the velocity in the real space 
will be given by
\begin{equation}
V=\frac{d x}{d t}=
\frac{\nu \partial_\chi A + \partial_\tau A}
{\nu \partial_\chi B + \partial_\tau B},
\end{equation}
which, by hypothesis, does not depend on $m$. 

This  is inconsistent with  (\ref{vxi}) where  a mass  dependence is
present and therefore $A$ and $B$ defined above do not exist. 

Let us point out that this is in agreement with the result in
\cite{noiagain}, where it is shown that it is not possible to define such map
and to keep the notion of invariant length scale. 

%%%%%%%%%%%%%%%%%%%%%%%%%%%%%%%%%%%%%%%%%%%%%%%%%%%%%%%%%%%%%%%%%%%%%%
%%%%%%%%%%%%%%%%%%%%%%%%%%%%%%%%%%%%%%%%%%%%%%%%%%%%%%%%%%%%%%%%%%%%%%
\section{Deformed derivatives}
%%%%%%%%%%%%%%%%%%%%%%%%%%%%%%%%%%%%%%%%%%%%%%%%%%%%%%%%%%%%%%%%%%%%%%
%%%%%%%%%%%%%%%%%%%%%%%%%%%%%%%%%%%%%%%%%%%%%%%%%%%%%%%%%%%%%%%%%%%%%%

In this last section we will consider another definition of velocity
as  was first discussed  by Lukierski  and Nowicki  in the  context of
$\kappa$-Poincar\'e formulation \cite{lukvel}. 

Let  us consider the  function $E(p)$,  that is  the solution  for the
energy  that comes from  the Casimir  defined in  (\ref{casdsr}).  The
velocity, as we  have been discussing up to now,  is the derivative of
such function with respect to $p$, that is
$$
V=\lim_{p_1 \to p_2}\frac{E(p_2)-E(p_1)}{p_2-p_1}.
$$
This definition involves the difference of the energy (evaluated in two 
different points) and a difference of
momenta, but this operation is only well defined through the use of composition
law of  four momenta \footnote{See for instance  G. Amelino-Camelia in
  {\em gr-qc/0309054}.}, that
is
\begin{eqnarray}
\label{diference}
\hat{\delta}p&=&p_a \hat{-}~p_b,\nonumber
\\
&=&F^{-1}[F[p_a] - F[p_b]],
\end{eqnarray}
with $F$ defined in (\ref{f}) for DSR1. 

Then, a DSR1 inspired definition of velocities is 
\begin{eqnarray}
\label{alvel}
\hat{V}&=&\lim_{p_1 \to p_2}\frac{E(p_2)\hat{-}E(p_1)}{p_2\hat{-}p_1},
\\
&=&\lim_{p_1 \to p_2}\frac{\hat{\delta}E}{\hat{\delta}p}.
\end{eqnarray}
That is, the comparison of the quantities are expressed in terms of
covariant differences.  

In order  to take  the limit we  use (\ref{diference})\footnote{Notice
  that as an  argument of $E$ through the  dispersion relation $p$ has
  to be considered as a numerical parameter and it is correct to write
  $p_2 =p_1+\delta$.}. The result is
\begin{eqnarray}
\label{diff2}
\hat{\delta}p^{\mu}&=&(p_2 \hat{-}~p_1)^\mu, \nonumber
\\
&=&(F^{-1})^{\mu}[F[p_1+\delta] - F[p_1]],\nonumber
\\
&=&(F^{-1})^{\mu}\left[\frac{\partial  F}{\partial E}~\vline_{p_1=p}\delta^0
  +\frac{\partial F}{\partial p}\vline_{p_1=p} \delta^1 \right],
\end{eqnarray}
where we have  restored the indexes explicitly (for  simplicity we are
in 1+1 dimensions, therefore $\mu =0,1$).

Using the Casimir relation
$$
C(E,p)=\cosh(\lambda               E)-\frac{\lambda^2}{2}p^2~e^{\lambda
  E}-\cosh(\lambda m),
$$
 namely the fact that free particles move on the orbits of $C(E,p)=0$ and the
variations we are considering must satisfy $\delta C(E,p)=0$
\begin{eqnarray}
\label{variac}
\delta^0&=&
-\left(\frac{\partial C}{\partial p}\right)
\left(\frac{\partial C}{\partial E}\right)^{-1}~\delta^1,
\nonumber\\
&=& \frac{dE}{dp}~\delta^1,\nonumber\\
&\equiv& V~\delta.
\end{eqnarray}

Now we can write the velocity of the particle as
\begin{equation}
\hat{V}=\lim_{\delta\to
  0}\frac{\left(F^{-1}\right)^0\left[\left(\partial_{p}F+
  V~\partial_{E}F\right)\delta\right]}{\left(F^{-1}\right)^1
  \left[\left(\partial_{p}
  F+V~\partial_{E}F\right)\delta
  \right]}.
\end{equation}

For DSR1, using the function $F$ defined in (\ref{f}) we get
\begin{equation}
\hat{V}=\frac{\lambda p  \cosh(\lambda m)}{{\cal C}^4_{\lambda m}(\lambda
  p)}\frac{\left[2{\cal C}^2_{\lambda m}(\lambda
  p) - r^2_{\lambda m}(\lambda p)\right]^2}{\left[2 {\cal C}^2_{\lambda m}(\lambda
  p)-r^2_{\lambda m}(\lambda p)({\cal C}^2_{\lambda m}(\lambda
  p)+1) \right]},
\end{equation}
with  $r_{\lambda m}(\lambda p)$  and ${\cal  C}^2_{\lambda m}(\lambda
p)$ defined in (\ref{vp}). 

The relation with the boost  parameter $\xi$ can be obtained by replacing
the  momentum  $p(\xi)$ given  in  (\ref{expboost}),  in the  previous
expression. The result is
\begin{equation}
\hat{V}(\xi)=\tanh(\xi).
\end{equation}
which is the standard relation between the boost parameter and the
velocity in the undeformed relativistic case.  However one can see that the
relation between the boost parameter and the momenta is
quite different compared with the undeformed case:
\begin{eqnarray}
E[\xi]&=&\frac{1}{\lambda}\ln\left(\cosh(\lambda m)+
\sinh(\lambda m)\gamma(\xi))\right),
\\
p[\xi]&=&\frac{1}{\lambda}
\frac{\gamma(\xi) \hat{V}(\xi)}{(\gamma(\xi) +\coth(\lambda m))},
\end{eqnarray}
with $\gamma(\xi)=(1-\hat{V}^2(\xi))^{-1/2}=\cosh(\xi)$. 

The fact that the relation between the velocity and the boost parameter
coincides with the standard of undeformed relativity --with the definition
of velocity adopted here-- might  suggest that there are no differences
between DSR proposal and the standard relativity. 

However, the interpretation of this velocity as the rate of
change of the space coordinate with respect to time is not guaranteed,
because  we do not  know which  deformation (if  any) of  the Poisson
brackets (simplectic  structure) between the coordinates  of the phase
space  ($x,p$) could  permit to  write this  deformed velocity  as the
result of a Hamilton equation.

Therefore, it would be premature to conclude that, with this definition
of velocity,  there are no  testable phenomenological consequences,
in  particular   for  the  propagation   of  light.   In   fact  these
consequences rely on finite time and distance measurements which are still
undefined, even in this approach.
%%%%%%%%%%%%%%%%%%%%%%%%%%%%%%%%%%%%%%%%%%%%%%%%%%%%%%%%%%%%%%%%%%%%%%
%%%%%%%%%%%%%%%%%%%%%%%%%%%%%%%%%%%%%%%%%%%%%%%%%%%%%%%%%%%%%%%%%%%%%%
\section{Discussion and conclusions}%%%%%%%%%%%%%%%%%%%%%%%%%%%%%%%%%%
%%%%%%%%%%%%%%%%%%%%%%%%%%%%%%%%%%%%%%%%%%%%%%%%%%%%%%%%%%%%%%%%%%%%%%
%%%%%%%%%%%%%%%%%%%%%%%%%%%%%%%%%%%%%%%%%%%%%%%%%%%%%%%%%%%%%%%%%%%%%%
In (undeformed) special relativity (SR) the assumption of constancy of
speed  of   light,  the  homogeneity   of  the  space   (linearity  of
transformations) together with the definition of velocity for massive
bodies  and   its  identification   with  frame  velocity   allows  to
constructively define (through {\it gedanken-}experiments) space-time.

Doubly special  relativity theories are constructed  in momentum space
by requiring  the existence of  an invariant momentum  (and/or energy)
scale which  assumes the meaning  of maximum momentum. The  reason for
this  is the  expectation that  Quantum Gravity  introduces  a minimum
(invariant) length  scale. However the construction  of the space-time
sector  is  still  in  its  infancy,  although  there  are  interesting
connections with the quantum deformed approaches.
We have in mind to try to repeat the undeformed SR construction, which
has  to pass  through  a  realistic definition  of  velocity for  both
massless and massive particles.

In this work we have discussed three different definitions of velocity
in a DSR1  scenario, as an approach to the  space time compatible with
these principles: {\em a}) the velocity as the derivative of the energy with
respect to the momentum, {\em b}) through the {\em classical} space-time 
approach and {\em c}) by a deformation of the concept of derivatives,
induced by the deformation of the addition law in the momentum space. 

Our first expression for the velocity is obtained by the natural
assumption
$$
v=\frac{dE}{dp},
$$ 
which is a consequence of two requirements, as was discussed in the
first part  of this  work: {\it i})  a Hamiltonian  (energy) function,
which is the generator of  time translations and {\it ii}) a canonical
simplectic structure.  Therefore, when  we adopt the above expression,
necessarily  we must  understand  $v$  as the  change  of the  spatial
coordinate with respect to time \footnote{We are calling \lq time \rq~
the canonical conjugate of the Hamiltonian.}, and then, we can extract
information from thought experiments by using the standard kinematics.

An  unavoidable   consequence  of  this  approach,   however,  is  the
dependence  of the  velocity on  the mass  of boosted  particles 
\cite{mignemi} (for photons, the  dependence of the
speed of  light on  the energy) which  implies, for example,  that two
particles of different mass at rest in some reference frame, will have
different velocities as seen  by another observer boosted with respect
to the first one. As a consequence, these two masses could interact (a
collision, for instance) for one observer while continue to be at rest
for the other.   The principle of relativity is  then violated.  Since
we require that DSR  respects the relativity principle, our assumption
for  the definition  of velocity  must  be discarded.  Another way  of
describing this  result is by  saying that reference frames  cannot be
unambiguously attached to massive particles or objects.

A possible way  out is to impose, as a  physical requirement, that all
bodies which are at rest according to one observer, move with the same
velocity for  any other  observer boosted respect  to the  first. This
solves the problem with the relativity principle, but it is not
compatible  with the  composition laws  of momenta  in DSR,  and makes
impossible to  associate in  a unique way  the boost parameter  with a
given reference frame.

Another  possibility, which  also  permits to  investigate a  possible
approach to DSR  in the space time as a non  linear realization of the
Lorentz Group, is to define the velocity in the {\it classical} space as
defined in Section IV and then to map it into the real space.

We have  shown that it  is not possible  to construct a  function that
maps a classical space-time, in which the Lorentz  group acts  linearly, 
into the real space-time.  We  have shown
that such function  would not be universal for  all particles because
it  will depend  on  the mass  of  each particle.  This  result is  in
agreement with a previous one obtained in \cite{noiagain}.

Finally, let  us comment our last  result.  We have
chosen a deformed definition for the derivative because, as we argued,
the difference (composition law) of energy and momentum has 
to be modified in order to be invariant under DSR. 
We have only one definition for the difference of
energy and momentum  that is compatible with DSR  principles, which is
\cite{noiagain}
$$
\delta p=p_b \hat{-}~p_a=p_b \hat{+}~S(p_a)=F^{-1}\left[F[p_b]-F[p_a]\right]
$$
where we have used the antipodal map $S[p]=F^{-1}[-F[p]]$. 

This law, inherited from the  composition law, should be the right one
that must appear in the  definition of derivatives. An example of this
kind  of  deformed derivatives  can  be  found  in the  definition  of
velocities in $\kappa$-Poincare scenario as given in
\cite{lukvel}. However there are  two  differences:
{\it i}) in DSR the composition law for the energy is modified while in
$\kappa$-Poincare it is the primitive one {\it ii}) the composition law of
momenta in DSR  is symmetric and therefore there  is only one possible
definition for derivative, instead in KP we have two possible choices
which, in fact, give rise to two possible velocities.

The result for DSR1 is a  definition of (three-)velocity which is that of
undeformed special relativity, while still maintaining an invariant
momentum scale; therefore we conclude that  DSR proposal
has, as  it must be,  two invariant scales,  namely $c$, the  speed of
light  and   $1/\lambda$,  the   maximum  momentum  attainable   for  a
particle.

The above defined velocity  also gives information about the space
time. If we follow the same argument  for the KP scenario, we will see that
the velocity found  in this case is in  agreement with the definition
of  velocity in  terms of  a Hamiltonian  and a  deformed simplectic
structure.  Then,  in very speculative  sense, we would say  that this
velocity for DSR could correspond to a deformed simplectic structure
and, at the end, to a space-time with a non trivial structure.

A posteriori  our results are  not unexpected since in  momentum space
the limit of infinitesimal incremental ratios makes sense even in presence of
an  (invariant) maximum  momentum, while  this is  clearly not  so for
space-time increments in  presence of a {\it minimum}  length. This is
at the base of the failure of attempts to construct directly a (continuous)
space-time   with  an   invariant   length  scale   as  described   in
\cite{noiagain}.

All our considerations have been discussed in the framework of DSR1, a
theory with limited momentum but unlimited energy. Similar discussions
can be carried out in different DSR flavors.

As a last, but very important remark, let us call the attention on the
fact that, since we do not  know the relation between  what we have
called {\em velocity}  and $dx/dt$, it is hard to  say whether the DSRs
can be  verified or disproved experimentally, for  instance by studying
the time of  flight of photons of different energies
from distant  sources; in particular we
are not allowed to  conclude that, since the deformed (three-)velocity
definition is identical with  undeformed special relativity, there are
no effects, since what one is really measuring are time and distances,
for which we have at the moment no definition.

\begin{acknowledgments}
We would like to thank G. Amelino-Camelia, J.  L.  Cort\'es, J.  Gamboa and
J.  Kowalski-Glikman for useful discussions on this
topic. Part of this work was developed during a Mini-Workshop
at LNGS in September 2004. F.M. thanks INFN for a postdoctoral fellowship. 
\end{acknowledgments}

\end{document}